\documentclass[12pt]{article}

\usepackage{graphicx}
\usepackage{enumerate}
\usepackage{natbib}

\usepackage{subfigure}
\usepackage{amssymb}
\usepackage{amsmath}
\usepackage{algorithm}
\usepackage{algorithmic}

\usepackage{booktabs}
\usepackage{multirow}
 \usepackage{hyperref}
 \usepackage{xcolor}
\usepackage{url} 

\newcommand{\blind}{1}

\usepackage{xr}
\makeatletter

\newcommand*{\addFileDependency}[1]{
\typeout{(#1)}
\@addtofilelist{#1}
\IfFileExists{#1}{}{\typeout{No file #1.}}
}\makeatother

\newcommand*{\myexternaldocument}[1]{%
\externaldocument[s-]{#1}%
\addFileDependency{#1.tex}%
\addFileDependency{#1.aux}%
}

\myexternaldocument{supp}

\begin{document}

\def\spacingset#1{\renewcommand{\baselinestretch}%
{#1}\small\normalsize} \spacingset{1}


\if1\blind
{
  \title{\bf LLOT: application of Laplacian Linear Optimal Transport in spatial transcriptome reconstruction}
  \author{Junhao Zhu\thanks{JZ and KZ  are  partially supported by CANSSI (Canadian Statistical Sciences Institute), Data Science Institute and Medicine by Design, University of Toronto}, Kevin Zhang\footnotemark[1], Dehan Kong\thanks{
    DK and ZZ acknowledge financial support from a 
    \textit{Catalyst Grant from Data Science Institute and Medicine by Design, University of Toronto. }}\hspace{.2cm}\\
    Department of Statistical Sciences, University of Toronto\\
    and \\
    Zhaolei Zhang \footnotemark[2] \\
    Donnelly Centre for Cellular and Biomolecular Research,\\ University of Toronto}
    \date{}
  \maketitle
} \fi

\if0\blind
{
  \bigskip
  \bigskip
  \bigskip
  \begin{center}
    {\LARGE\bf LLOT: application of Laplacian Linear Optimal Transport in spatial transcriptome reconstruction}
\end{center}
  \medskip
} \fi

\bigskip
\begin{abstract}
Single-cell RNA sequencing (scRNA-seq) allows transcriptional profiling, and cell-type annotation of individual cells. However, sample preparation in typical scRNA-seq experiments often homogenizes the samples, thus spatial locations of individual cells are often lost. Although spatial transcriptomic techniques, such as in situ hybridization (ISH) or Slide-seq, can be used to measure gene expression in specific locations in samples, it remains a challenge to measure or infer expression level for every gene at a single-cell resolution in every location in tissues. Existing computational methods show promise in reconstructing these missing data by integrating scRNA-seq data with spatial expression data such as those obtained from spatial transcriptomics. Here we describe Laplacian Linear Optimal Transport (LLOT), an interpretable method to integrate single-cell and spatial transcriptomics data to reconstruct missing information at a whole-genome and single-cell resolution. LLOT iteratively corrects platform effects and employs Laplacian Optimal Transport to decompose each  spot in spatial transcriptomics data into a spatially-smooth probabilistic mixture of single cells. We benchmarked LLOT against several other methods on datasets of \textit{Drosophila} embryo, mouse cerebellum and synthetic datasets generated by scDesign3 in the paper, and another three datasets in the supplementary. The results showed that LLOT consistently outperformed others in reconstructing spatial expressions. 
\end{abstract}

\noindent%
{\it Keywords:} Data Integration, Linear Map, Laplacian Optimal Transport, Spatial Expressions, scRNA-seq
\vfill

\newpage
\spacingset{1.9} 
\section{Introduction}\label{sec1}
Animal tissues have defined and precise 3-D structures, typically consisting of many different cell types. To understand the function and developmental processes of these tissues, it is crucial to identify both the types of cells and the location of these cells \citep{halpern2018paired}. 
Recent advances in single-cell RNA sequencing (scRNA-seq) have allowed transcriptional measurement and identification of cells at single-cell resolution, as well as determining the composition of cells in complex tissues. However, sample preparation prior to sequencing typically requires dissociation and homogenization of tissues, thus the spatial context of individual cells is often lost \citep{zhuang2021spatially}. One way to overcome such obstacles is to select a subset of “landmark” genes and use image-based technology such as in situ hybridization (ISH) to measure the expression level of these landmark genes and then infer the identity of the cells. One limitation of ISH is it can only measure a small number of genes, e.g., only 84 genes in the \textit{Drosophila} embryos data \citep{okochi2021model}. More recent technologies such as \citep{rodriques2019slide} can provide genome-wide expression at a high spatial resolution, but each spot in the tissue often contains many cells of heterogeneous types \citep{cable2022robust}.
Despite advances in several computational methods, whole-genome spatial reconstruction at a single-cell resolution is still challenging.
In the following, we formally define this problem, give an overview of existing approaches, and describe the rationale of our proposed method. 

In the realm of whole-genome spatial gene expression, there are two inter-connected questions to be answered:
1. What is the gene expression level of all the cells in aggregate in this location? 
2. For cells of unknown origin, can we accurately assign them to specific locations in the tissue based on their expression profiles? 
To achieve these goals, one typically requires some prior knowledge of spatial expression information of some landmark genes (e.g., as measured by ISH \citep{Distmap})  or alternatively from genome-scale experiments (e.g., as measured by Slide-seq \citep{rodriques2019slide}).

Several computational methods have been proposed to tackle these problems \citep{Distmap,stuart2019comprehensive,liger,novaspa,okochi2021model}. The common theme of these methods is to integrate the scRNA-seq data, which provides genome-wide expression profiles of individual cells, with existing spatial expression data of selected landmark genes as a reference. 
 Distmap first binarizes expression levels of the same set of  marker genes in both spatial transcriptomics data and scRNA-seq data according to their quantiles and calculates Mathews correlation coefficients for every bin to assign single cells to specific locations \citep{Distmap}. However, this method is prone to low accuracy since binarization can lead to a loss of information.
As an improvement over binarization comparison,  several other methods apply  dimensionality-reduction on common marker genes and compare scRNA-seq and spatial transcriptomics data in the low dimensional space. For example, Seurat \citep{stuart2019comprehensive}  uses  canonical correlation analysis \citep{hardoon2004canonical}  and Liger \citep{liger} applies  integrative non-negative matrix factorization \citep{yang2016non}. These two methods then utilize the nearest-neighbor method  to link spatial transcriptomics data with scRNA-seq data; however, these two methods have the limitation of high computational cost in searching nearest neighbors. Perler \citep{okochi2021model} applies Gaussian mixture model to integrate datasets, while it assumed all the input data is normally distributed and 
suffered  from large computational cost due to high-precision evaluation of 
Gaussian density in high dimensional space. 
In addition, these methods do not explicitly utilize the physical location of spots in the spatial transcriptomics data, and only use the information on gene expressions. 
In contrast to these aforementioned methods, researchers also developed methods based on the principle of optimal-transport  such as novoSpaRc \citep{novaspa},
which explicitly uses the spatial information of the spots.
These methods create a probabilistic mapping between scRNA-seq data and spatial transcriptomics data by minimizing the Gromov-Wasserstein distance between the gene expression space and the  physical space in tissues. Despite promising results, these methods assume that gene expression spaces and  physical spaces of tissues are homogeneous,
and do not account for platform effects.
As demonstrated in a study by \cite{cable2022robust},
these shortcomings can result in  under-performing of these methods due to the fact that biological signals can be masked by systematic technical variability \citep{leek2010tackling}.
Overall, these methods perform reasonably well in predicting spatial expressions. 
However, they often do not model differences in gene expressions across distinct datasets or do not consider the underlying spatial topology of the tissue.

To address these issues, we herein describe a new probabilistic framework called \textit{Laplacian Linear Optimal Transport} (LLOT).  Compared to existing methods, LLOT offers two key innovations. 
First, it explicitly models the platform effect by using a linear map to compare and calibrate the expression levels of common genes in the spatial expression dataset and in the scRNA-seq dataset. 
 The advantage  of the linear map is  that it is both interpretable and computationally efficient. Secondly, LLOT uses a Laplacian optimal transport approach to effectively model the relationship between the spatial distance between cells and their expression similarities.
 The advantage of the Laplacian optimal transport is that  it does not assume a  homogeneity between cell-cell physical distance and expression similarity and helps capture  complex spatial structures of tissues.

We applied LLOT method to predict spatial expressions of three pairs of spatial transcriptomics datasets and scRNA-seq datasets, a \textit{Drosophila}  embryo dataset\citep{Distmap}, a  mouse cerebellum dataset \citep{cable2022robust},  a processed mouse hippocampus dataset \citep{li2022benchmarking}, and synthetic datasets simulated by scDesign3 \citep{song2024scdesign3},
and benchmarked the prediction performance with four state-of-the-art methods, novoSpaRc \citep{novaspa},  Seurat \citep{stuart2019comprehensive} and Liger \citep{liger}.  Our benchmark results showed that LLOT is consistently among the better methods or
outperforms these other methods in spatial expression predictions, judged by Pearson  correlation coefficients (PCC) between predicted spatial expressions versus ground truth. 
We also demonstrated that LLOT can effectively recover segmentation-involved expression patterns  in \textit{Drosophila}  embryo. In the supplementary  materials, we  extended LLOT to reconstruct layers in human dorsolateral prefrontal cortex data \citep{maynard2021transcriptome} as a classification task and to perform cell-type deconvolution on a simulated benchmarking mouse visual cortex dataset \citep{li2022benchmarking}, and benchmarked against other state-of-the-art methods. These results demonstrated that LLOT, as an interpretable framework, can effectively predict spatial expressions and infer cell locations. 



The remainder of our paper is organized as follows. In Section \ref{sec:bg}, we introduce the background and notation. In Section \ref{sec:method}, we introduce the structures of spatial expression and scRNA-seq data, the tasks we aim to solve, and describes our LLOT method. In Section \ref{sec:res}, we provide benchmark study of gene reconstruction accuracy  on a \textit{Drosophila} embryo data and a mouse cerebellum data against several state-of-the-art methods, and analyze the  results of LLOT in \textit{Drosophila} embryo. We then conduct numerical simulation study in Section \ref{sec:simulation}. 
We end with a discussion in Section \ref{sec:dis}. 


\begin{figure}[]
\centering  
\includegraphics[width=0.9\textwidth]{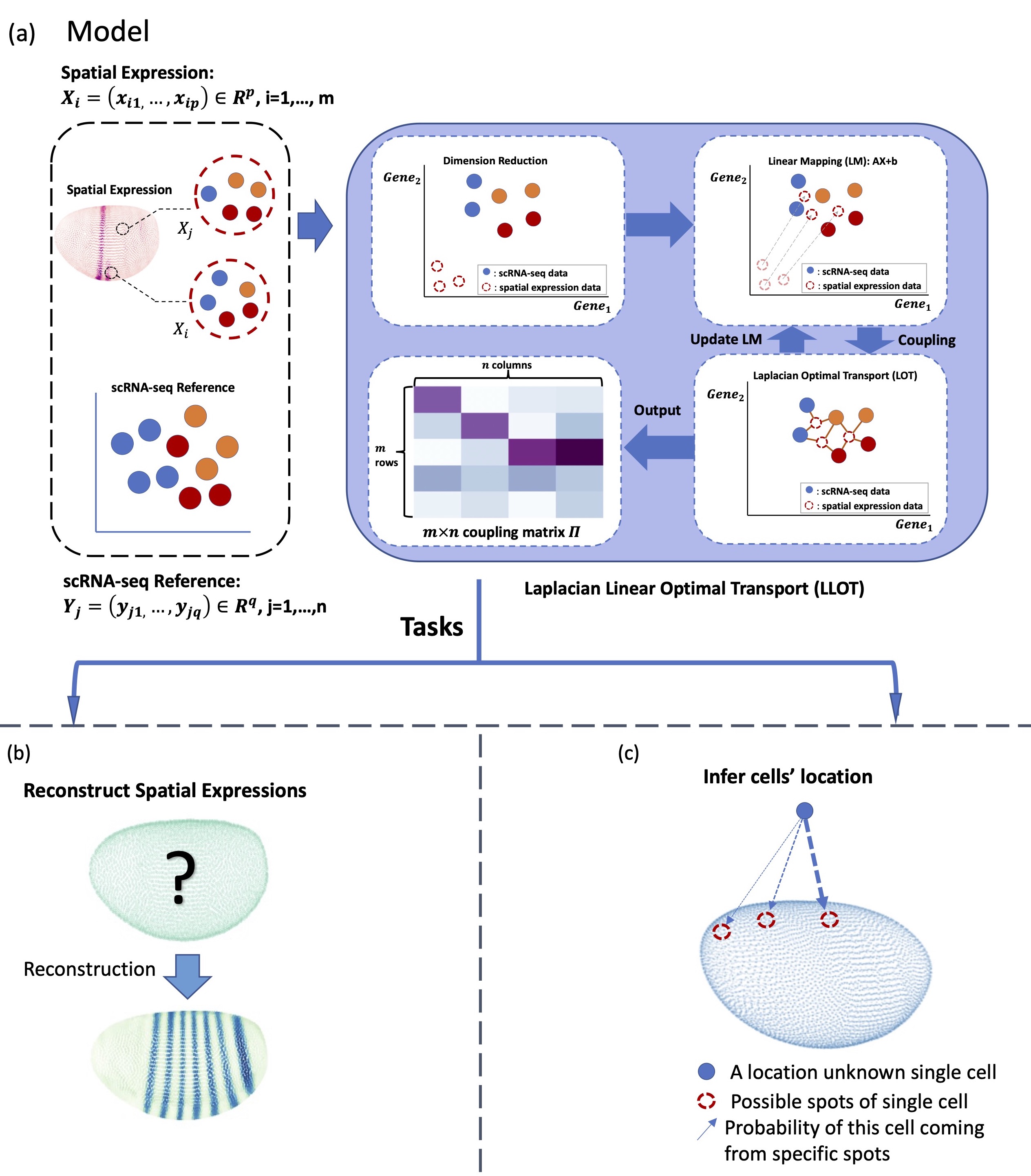}  
\vspace{-20pt}
\caption{\textbf{Schematic Overview of LLOT.}
\textbf{(a)} Left: Input data are a spatial expression dataset (ISH or Slide-seq) with unknown cells and a scRNA-seq  dataset without spatial locations.  A spatial expression dataset has $m$ spots measurement, where each spot has $p$ genes measured; a scRNA-seq dataset contains $n$ individual cells with $q$ genes measured. 
Right: LLOT  implements multiple iterations of platform effect estimation and coupling estimation. The output of LLOT is an $m \times n$ coupling matrix $\Pi$, representing single-cell composition of each spot. Elements $\Pi_{i,j}$
indicates the probability that single cell $j$ is in the spot $i$.  \textbf{(b)}. Reconstruction of spatial gene expression.  \textbf{(c)}. Inferring cells' locations..}
\label{Fig.main}
\end{figure}

\section{Background}\label{sec:bg}
Throughout this paper, we define ``spatial expression dataset'' as expression levels of selected  genes in known spatial locations, and define ``scRNA-seq dataset'' as a whole-genome single-cell expression dataset of a population of cells of unknown spatial locations ($\textbf{Fig.}$ \ref{Fig.main}a).  

\begin{figure}
\centering  
\subfigure[ISH data and scRNA-seq data]{
\includegraphics[width=0.95\textwidth]{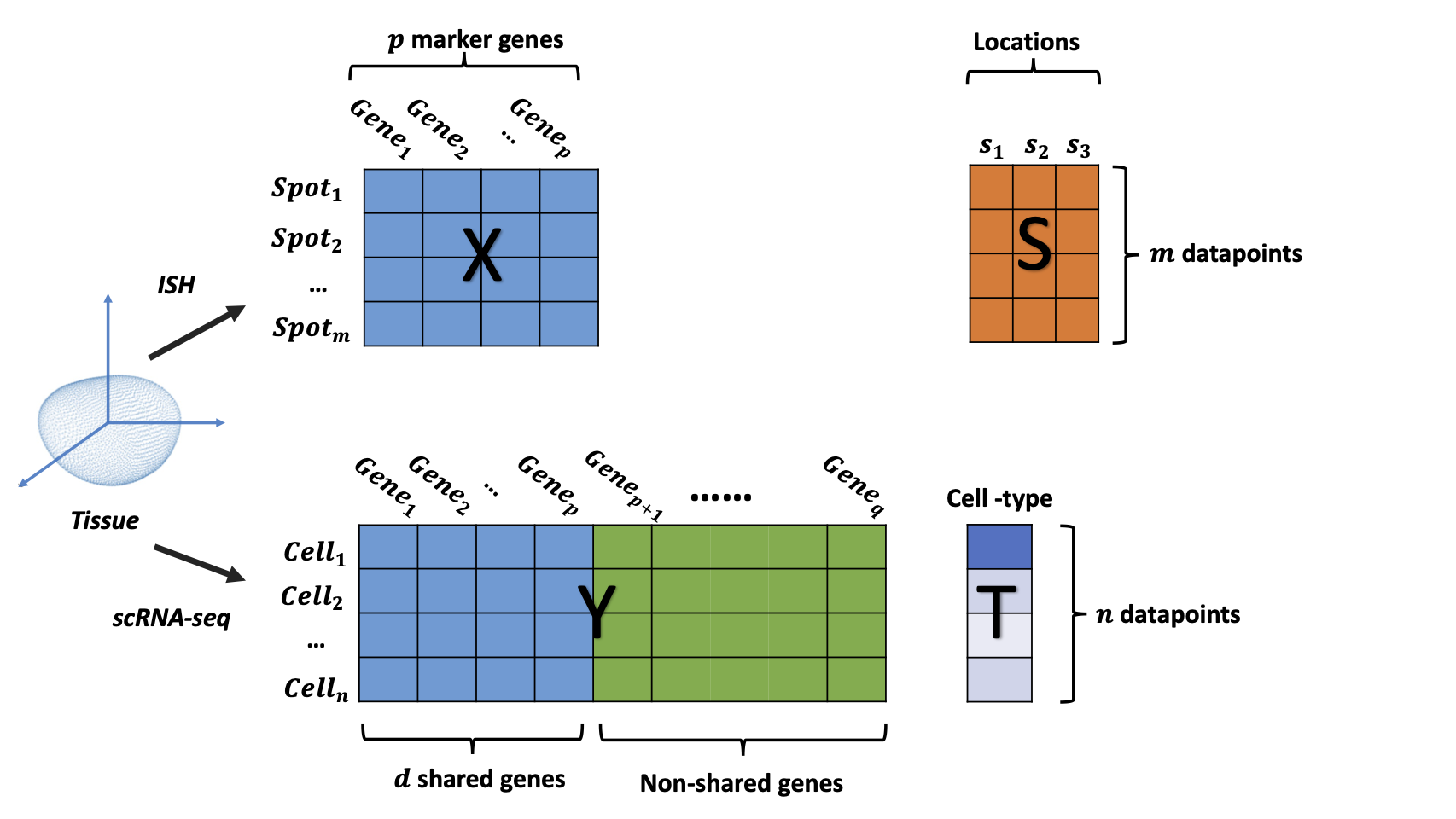}}
\subfigure[Slide-seq data and scRNA-seq data]{
\includegraphics[width=0.95\textwidth]{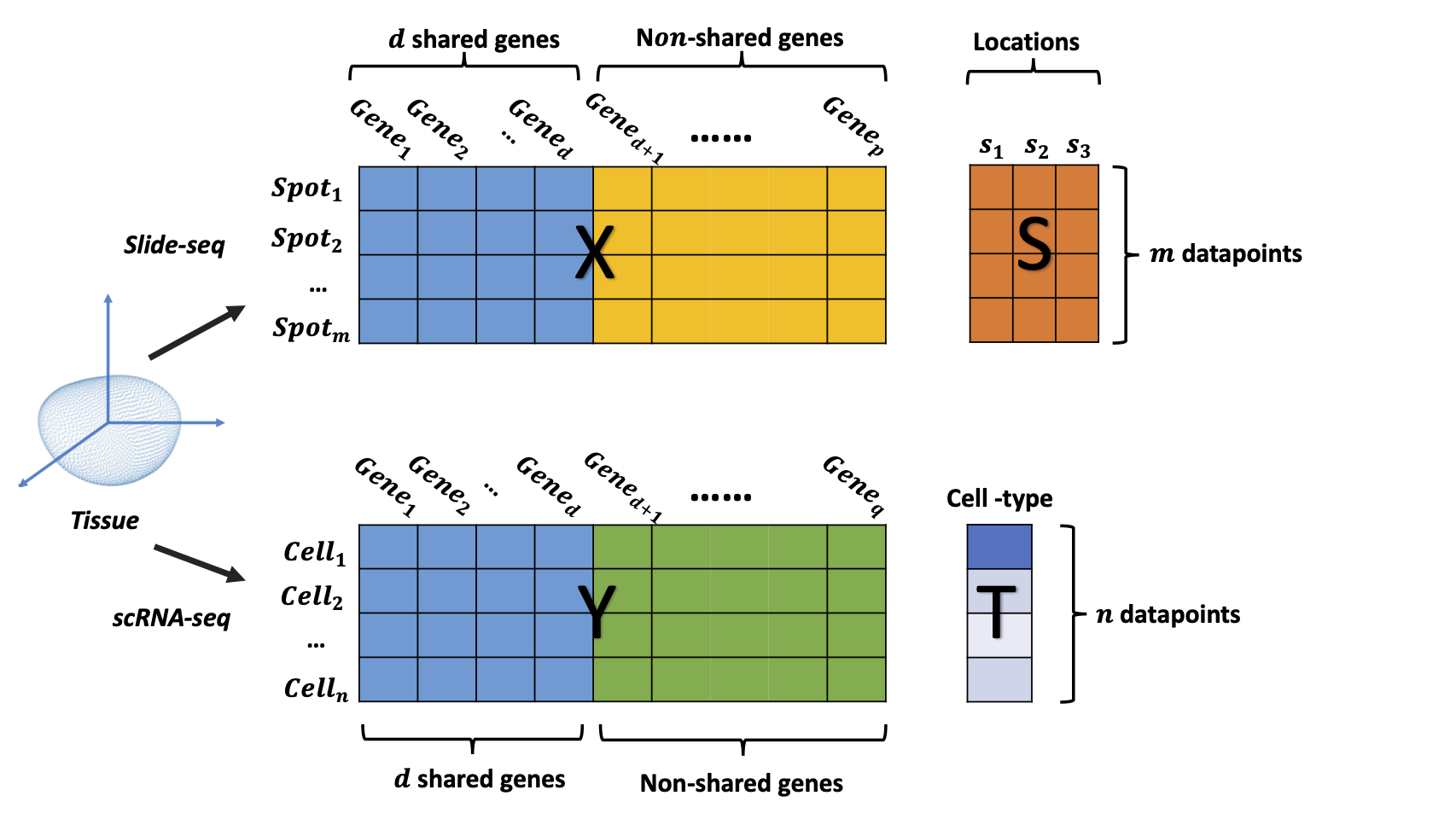}}
\caption{\textbf{Overview of Data Structure.}  \textbf{(a).}  An overview of data structures of ISH data and scRNA-seq data ($p=d$). \textbf{(b).} An overview of data structures of Slide-seq data and scRNA-seq data. 
The blue matrices represent the genes shared by both datasets, while the yellow matrix and the green matrix represent genes that are unique in each dataset.
}
\label{DataStructure}
\end{figure}

\subsection{Notation}
To present our problem and method, we first introduce notation and describe the data structure. We refer to the spatial transcriptomic dataset as $\mathcal{X}$ and the scRNA-seq data as $\mathcal{Y}$. In $\mathcal{X}=\{\textbf{X},\textbf{S}\}$, we have measurements from $m$ spots whose locations in the tissue are known. Here $\textbf{X}\in\mathbb{R}^{m\times p}$ is a matrix of size $m\times p$ representing the expression profiles of $p$   genes in the $m$ spots, and $\textbf{S}\in \mathbb{R}^{m\times 2}$ is a matrix of size $m\times 2$ indicating the corresponding  2-D locations of measured spots in the tissue (which can be generalized to 3-D locations). There are two major types of spatial transcriptomics data \citep{li2022benchmarking}. The first type is  generated by image-based methods such as  in situ hybridization (ISH)\citep{Distmap}, which can detect spatial expressions with high resolution and accuracy, but is limited in the total number of RNA transcripts that they can detect. 
The second type is generated by sequencing-based methods such as Slide-seq \citep{rodriques2019slide}, which has access to genome-wide spatial expressions for all spots in tissues, but each spot may contain multiple cells.
To be specific, let $X_i=(x_{i,1},\cdots,x_{i,p})$ be the $i^{th}$ row of the matrix $\mathbf{X}$ and $S_i= (s_{i,1},s_{i,2})$ be the $i^{th}$ row of the matrix $\mathbf{S}$, where $x_{i,k}$ represents the expression level of the $k^{th}$ gene measured by the spatial transcriptomics in the spot $i$ and   $(s_{i,1},s_{i,2})$ is the 2D-coordinate of the spot $i$ in the tissue.  
To give a better understanding, we present the data structure of these two types of spatial expression datasets in \textbf{Fig.}\ref{DataStructure}. 
For the scRNA-seq data $\mathcal{Y}=\{\textbf{Y}\}$, we have genome-wide expression profiles 
of $n$ single cells without spatial information. Here $\textbf{Y}\in\mathbb{R}^{n\times q}$ is a numeric matrix of size $n\times q$ representing expression profiles of $q$ genes in $n$ single cells (e.g., $n =1,207$ and $q=8,924$  in the \textit{Drosophila} embryo scRNA-seq data with $11$ major cell types). To be specific, we denote $Y_j=(y_{j,1},\cdots,y_{j,q})$  the $j^{th}$ row of the matrix $\mathbf{Y}$ and $T_j$  the $j^{th}$ element  of the categorical vector $\mathbf{T}$, 
where $y_{j,k}$ represent the expression level of the $k^{th}$ gene of the $j^{th}$ single cell in the scRNA-seq and   $T_j$ indicates the cell-type label of the $j$th single cell for $ j=1, \ldots, n$. 
For the spatial transcriptomic data $\mathcal{X}$ and the scRNA-seq data $\mathcal{Y}$, they share $d$ common genes  (e.g., $d=84$ in the \textit{Drosophila} embryo ISH dataset and the  \textit{Drosophila} embryo scRNA-seq dataset, and $d=21,056$ in the mouse cerebellum Slide-seq and scRNA-seq dataset).
For simplicity, we assume that the $d$ genes shared by both datasets are represented by column $1$ to column $d$, and use $X_{i,1:d}=(x_{i,1},\cdots,x_{i,d})$ to represent the aggregated gene expression levels in the spot $i$ and  $Y_{j,1:d}=(y_{j,1},\cdots,y_{j,d})$ to represent the expression profile of the $j^{th}$ single cell for the $d$ shared genes between two datasets. 

It is important to infer whole-genome spatial gene expression at a  single-cell resolution, which requires us 
to solve the following two tasks as  shown in \textbf{Fig.}\ref{Fig.main}b-c:
\begin{itemize}
    \item How to reconstruct spatial expressions of genes at a single-cell resolution  that are not observed in the spatial transcriptomic dataset?
    \item  Given a specific single cell with a genome-wide expression profile, how to infer the possible locations of this cell in the tissue?
\end{itemize}
\section{Method}\label{sec:method}
Given the data, the above problems  can be translated to  two probability problems as follows:  1.
    Given the gene measurements $X_i$ of the spot $i$ in the spatial transcriptomics data, assume we can measure a single cell's genome-wide expression profile at this particular spot using scRNA-seq, 
    what is the conditional probability distribution of the genome-wide gene expression levels of this single cell?
  2. Given a single cell data point with genome-wide expression profile $Y_j$, what is the conditional probability that this cell originates from each spot?

 The conditional probability distributions above are not directly observable in practice. For example, in the ISH data, we can only measure a limited number of marker genes, such as $84$ genes in our \textit{Drosophila} embryo data, which is not representative of the genome-wide expression. Similarly, in the Slide-seq data, a single spot may contain multiple cells, making it difficult to obtain single-cell resolution. Furthermore, in the scRNA-seq data, spatial information about cells is lost since tissues are dissociated
before cells are sequenced. 

In this paper, we utilize a coupling matrix  to solve these problems. In mathematics, conditional probability distributions are specified by a coupling matrix (or equivalently a joint distribution) which is not directly observable in practice. A coupling matrix $\Pi\in\mathbb{R}_+^{m\times n}$ is a matrix  of size $m\times n $ that assigns probabilities to each spot-cell pair (spot $i$, cell $j$). In particular, if we could simultaneously measure the expression profile of the spot $i$ using Slide-seq or ISH and sequence a single cell located at the same spot using scRNA-seq, then the entry $\Pi_{i,j}$ in the coupling matrix represents the joint probability that the measured expression profile of spot $i$ is $X_i$ and the expression profile of the single cell is $Y_j$. In our scenario, the element $\Pi_{i,j}$ in the coupling matrix $\Pi$ scores the confidence that the spot $i$ in the spatial transcriptomics data contains the $j^{th}$ single cell in the scRNA-seq dataset. We could solve two tasks in  
Figure \ref{Fig.main} by normalizing the corresponding rows and columns of the coupling matrix; the details are given in the remainder of this section. 
We aim to infer an optimal coupling matrix such that it can best represent our observed data. In the remainder of this section, we will delve into the concept of the coupling matrix and demonstrate how it can be utilized to solve biological problems.

\subsection{Coupling Matrix and Optimal Transport.} 
 We aim to tackle the challenge of integrating two datasets: the spatial transcriptomic dataset containing $m$ spots $\{(X_i,S_i)\}_{i=1}^m$, which lacks single-cell genome-wide resolution, and the scRNA-seq dataset containing a collection of $n$ single-cell genomic expression profiles and cell-type labels $\{(Y_j,T_j)\}_{j=1}^n$ without  spatial coordinates. Although we cannot observe the coupling matrix directly, i.e., the joint distribution of the two datasets, our goal is to infer an optimal coupling matrix that best fits our data. 
 The problem of finding a coupling matrix  is referred to as the
 optimal transport (OT) framework. 

The estimation of an optimal coupling matrix is equivalent to   an optimal transport (OT) problem, which roots in  management science and the study of diffusing particles \citep{AIHP_1932, kantorovich1942translocation}. Recently, OT has been adopted in the study of single-cell lineage tracing and trajectory inference \citep{forrow2021lineageot}, and data integration of spatial transcriptomics and single-cell RNA-seq data \citep{novaspa}. Mathematically, the OT problem is formulated as a minimization problem:
\begin{equation}\label{OT0}
\begin{aligned}
&\min_{\Pi} \sum_{i,j}\Pi_{i,j}\|X_{i,1:d}-Y_{j,1:d}\|^2_2.
\\&\text{subject to}~
    \Pi_{i,j}\geq 0,~ \sum_{j}\Pi_{i,j}=\frac{1}{m}~\text{and}~\sum_{i}\Pi_{i,j} =\frac{1}{n},
\end{aligned}
\end{equation}
The above problem finds a coupling matrix that optimizes the average distance between two datasets.  It has appealing theoretical properties and can be solved using efficient algorithms \citep{peyre2019computational}. 
The constraints $\sum_{j}\Pi_{i,j}=\frac{1}{m}$ and $\sum_{i}\Pi_{i,j}=\frac{1}{n}$ indicate that the coupling $\Pi$ should be consistent with the marginal distributions of the observed data. 
To be specific, the marginal distribution of the spatial transcriptomics dataset is the empirical distribution of spatial gene expressions, attributing $1/m$ probability mass to each data point in the spatial expression dataset.  Thus the first constraint $\sum_{j}\Pi_{i,j}=\frac{1}{m}$ requires the assignment of single-cell data to spots should align with the empirical distribution of  the spatial gene expressions. 
The marginal distribution of scRNA-seq dataset, which is the empirical distribution of gene expression profiles of single cells (i.e., assigning $1/n$ probability mass to each scRNA-seq data point), represents the proportion of different cellular expression profiles and cell types in the tissue. Thus, the second constraint indicates that the predicted genome-wide spatial expressions or the estimated cell-type mixtures should be consistent with the exact cellular composition  measured by the scRNA-seq. The marginal distributions can also be modified to reflect prior knowledge or cell quality considerations. For instance, lower probabilities can be assigned to low-quality cells, while larger probabilities could be given to high-quality cells, thereby integrating quality considerations into the data analysis process.
The minimization in \eqref{OT0} involves assigning high probabilities to spot-cell pairs with similar gene expression profiles and low probabilities to those with large differences. However, this approach does not contain the platform effects and does not explicitly use the spatial information provided by the spatial transcriptomics data, as it only considers gene expression profiles. In the following subsection, we will explain how these issues can be addressed by our proposed method called  Laplacian Linear Optimal Transport (LLOT).

\subsection{Laplacian Linear Optimal Transport.}
The original OT in Eq.\eqref{OT0} does not account for  the platform effects which arise from differences in capture rates between the different platforms. This effect can be caused by genes being more sensitive to certain platforms than others, resulting in scale and mean differences even for shared genes. To account for this, we make the assumption that platform effects consist of two components: scale and average effects. This assumption, which is also shown in \cite{cable2022robust}, is easy to interpret and enables us to integrate a linear map into the OT objective (3) to account for gene-specific platform variations.
Let $\boldsymbol{a} = (a_1,\cdots,a_d)$ and  $\boldsymbol{b} = (b_1,\cdots,b_d)$  be two $d$-dimensional vectors, where $a_k$ represents the platform effects of scale in the $k^{th}$ common gene, and $b_k$ represents the platform effects of average in the $k^{th}$ common gene.  For simplicity,  let $\boldsymbol{a}\cdot{X_i} +\boldsymbol{b}$ be the linear map $(a_1 x_{i,1}+b_1,\cdots, a_k x_{i,k}+b_k,\cdots,a_d x_{i,d}+b_d)$. We introduce an OT with a linear map accounting for platform effects as follows:
\begin{equation}\label{lot}
\begin{aligned}
&\min_{\Pi,\boldsymbol{a},\boldsymbol{b}} \sum_{i,j}\Pi_{i,j}\|\boldsymbol{a}\cdot{X_i} +\boldsymbol{b}-Y_j\|^2_2.
\\&\text{subject to}~a_k>0,~b_k>0,~
    \Pi_{i,j}\geq 0,~ \sum_{j}\Pi_{i,j}=\frac{1}{m}~\text{and}~\sum_{i}\Pi_{i,j} =\frac{1}{n}.
\end{aligned}
\end{equation}  
Mathematically, to effectively minimize the objective function, it is crucial to perform an accurate estimation of the platform effects between the two datasets and to assign high scores $\Pi_{i,j}$ to the spot $i$ and the single cell $j$ if they exhibit similar gene expression profiles after correcting for the platform effects. 

Solving equation \eqref{lot} is not computationally efficient because 
OT is a constrained linear programming problem and the computational cost of approximating its numerical solution is $O((m+n)^3\log(m+n))$, which is not practicable in large-scale datasets \citep{cuturi2013sinkhorn}.
To improve the computational efficiency, we add an entropy regularization $\sum_{i,j}\Pi_{i,j}\log\Pi_{i,j}$ to the objective function in Eq.\eqref{lot}. The entropic regularization will significantly speed up the computation by making the objective function strongly convex, and the optimal solution can be efficiently approximated using matrix scaling algorithms \citep{cuturi2013sinkhorn}. For comprehensive details of the entropic regularization in optimal transport, we refer readers to  \cite{cuturi2013sinkhorn}.

Another problem is that the optimal coupling $ \Pi$ obtained by equation \eqref{lot} only takes into account the gene information, but  does not explicitly utilize the spatial context of tissues or the physical location information of spatial transcriptomics data. Specifically, if  $S_i$ and $S_j$ are closed in the physical space, then cells in the spot $i$ should have similar gene-expression profiles to  cells at spot $j$. Based on this fact, we can naturally introduce a Laplacian regularization to the objective function \eqref{lot} to incorporate the  information contained in spatial coordinates. The Laplacian regularization has been proved to has the ability to effectively capture the complex manifold and topology structure in the task of manifold learning \citep{lapeigen}. 

In the subsequent sections, we delve deeper into the specifics of the Laplacian regularization, explaining how it is utilized in our computational framework.
Let $ \boldsymbol{D}$ be a K-NN connectivity matrix of the spatial coordinates in the spatial transcriptomics dataset: 
\begin{equation}
    \boldsymbol{D}_{ij} =\left\{\begin{aligned}
&1\qquad &\text{if}~S_j~\text{is a K-nearest-neighbor of}~ S_i ~\text{or vice versa}\\&0&\qquad \text{otherwise}.
    \end{aligned}\right.
\end{equation}
\smallskip
 Based on the K-NN matrix, we introduce the Laplacian Linear Optimal Transport (LLOT) as follows:
 \begin{equation}\label{OTb}
\begin{aligned}(\hat{\Pi},\boldsymbol{\hat{a}},\boldsymbol{\hat{b}}) =&\arg\min_{\Pi,\boldsymbol{a},\boldsymbol{b}} \left(\sum_{i=1}^m\sum_{j=1}^n\Pi_{i,j}\|\boldsymbol{a}\cdot{X_i} +\boldsymbol{b}-Y_j\|_2^2+ \right.\\&\qquad\left.\lambda_1\sum_{i=1}^m\sum_{j=1}^n\Pi_{i,j}\log\Pi_{i,j}+\lambda_2\sum_{i=1}^m\sum_{j=1}^m \boldsymbol{D}_{ij}\sum_{k=1}^n(\Pi_{i,k}-\Pi_{j,k})^2\right),
      \\&\text{subject to}~a_k>0,~b_k>0,~
    \Pi_{i,j}\geq 0,~ \sum_{j}\Pi_{i,j}=\frac{1}{m}~\text{and}~\sum_{i}\Pi_{i,j} =\frac{1}{n},
\end{aligned}
\end{equation}
where $\lambda_1,~\lambda_2>0$ are regularization parameters. The entropic term 
$\sum_{j=1}^n\Pi_{i,j}\log\Pi_{i,j}$ is to improve the computational efficiency \citep{cuturi2013sinkhorn}.
The term $\sum_{i=1}^m\sum_{j=1}^m \boldsymbol{D}_{ij}\sum_{k=1}^n(\Pi_{i,k}-\Pi_{j,k})^2$ is called Laplacian regularization. The Laplacian regularization penalizes the squared difference between the rows of the coupling matrix corresponding to spots that are closed in the tissue, 
 encouraging the coupling matrix to be locally smooth with respect to the physical distances of the spots. This promotes the sharing of information between neighboring spots and enhances the spatial coherence of the estimated coupling matrix.

Denote $\boldsymbol{L}$ the Laplacian matrix of $\boldsymbol{D}$ \citep{flamary2016optimal}.  
Then \eqref{OTb} can be re-written as the following Laplacian regularized matrix form:
\begin{equation}\label{LLOT1}
\begin{aligned}(\hat{\Pi},\boldsymbol{\hat{a}},\boldsymbol{\hat{b}}) =&\arg\min_{\Pi,\boldsymbol{a},\boldsymbol{b}} \left(\sum_{i=1}^m\sum_{j=1}^n\Pi_{i,j}\|\boldsymbol{a}\cdot{X_i} +\boldsymbol{b}-Y_j\|_2^2+ \right.\\&\qquad\left.\lambda_1\sum_{i=1}^m\sum_{j=1}^n\Pi_{i,j}\log\Pi_{i,j}+\lambda_2 \text{tr}( \Pi^\top L\Pi)
\right),
      \\&\text{subject to}~a_k>0,~b_k>0,~
    \Pi_{i,j}\geq 0,~ \sum_{j}\Pi_{i,j}=\frac{1}{m}~\text{and}~\sum_{i}\Pi_{i,j} =\frac{1}{n}.
\end{aligned}
\end{equation}

\subsection{Estimation of parameters}\label{sec:method.est}
To tackle the optimization problem in Eq.\eqref{LLOT1}, we adopt a combination of the generalized conditional gradient (GCG) algorithm proposed by \citet{flamary2016optimal} and the coordinate descent algorithm \citep{wright1999numerical}. Our goal is to iteratively estimate the parameters $(\boldsymbol{a},\boldsymbol{b},\Pi)$. For fixed tuning parameters $ \lambda_1$ and $ \lambda_2$, we propose the following algorithm to solve the optimization problem Eq.\eqref{LLOT1}. 


\begin{enumerate}\label{alg}
    \item Input number of iterations $N$, learning rate $\gamma_t$ , batch size $B$, the initial guess of $(\boldsymbol{a^0}, \boldsymbol{b^0},\Pi^0)$ and index set  $\mathcal{I}=\{(i,j):~i=1,2,\cdots,m,~j=1,2,\cdots,n \}$. 
    \item For $t$ in $0:N$, repeat the following procedure:
    \begin{enumerate}
        \item Compute the squared distance matrix $C\in\mathbb{R}^{m\times n}$, where $C^t_{i,j} = \| \boldsymbol{a^t}\cdot{X_i} +\boldsymbol{b^t}-Y_j \|_2^2$ 
        and 
        $G^t= 2\lambda_2 \Pi^{t}L+C^t$.
        \item Solve the entropic optimal transport problem with cost matrix $G$ and entropy parameter $\lambda_1$, and the solution is denoted by $M$. Then we update the coupling matrix by $\Pi^{t+1} = \gamma_t M + (1-\gamma_t)\Pi^t$.
        \item \label{WLR} For $k$ from $1$ to $d$, perform weighted simple linear regression on $\{(x_i, y_j) : (i, j) \in \mathcal{I}\}$, using weights represented by $\Pi^{t+1}$, and denote the estimate as $(\alpha_k, \beta_k)$.
\item[$(\text{c}^\prime)$] \label{SLR} Alternatively, perform the accelerated stochastic linear regression as follows:
    \begin{enumerate}
        \item Randomly draw $B$ pairs of indices $(i,j) \in \mathcal{I}$ from the joint distribution represented by the coupling matrix $\Pi^{t+1}$, and denote the sampled pairs as $\mathcal{D}$.
        \item For $k$ from $1$ to $d$, perform simple linear regression on $\{(x_{i,k}, y_{j,k}) : (i,j) \in \mathcal{D}\}$, and denote the estimate as $(\alpha_k, \beta_k)$.
    \end{enumerate}
      \item Update parameters $(\boldsymbol{a}^{t+1},\boldsymbol{b}^{t+1})$ by $a^{t+1}_k = \gamma_t \alpha_k + (1-\gamma_t)a^t_k$ and 
        $b^{t+1}_k = \gamma_t \beta_k + (1-\gamma_t)b^t_k$.     
    \end{enumerate}
    \item Output $(\boldsymbol{a}^{N+1},\boldsymbol{b}^{N+1}, \Pi^{N+1})$ as the estimate $(\hat{\Pi},\boldsymbol{\hat{a}},\boldsymbol{\hat{b}})$.
\end{enumerate}
In practice,  the initial value of $a_k$ is set to be $\frac{\sigma_k(\mathbf{Y})}{\sigma_k(\mathbf{X})}$, where $\sigma_k(\mathbf{X})$ and $\sigma_k(\mathbf{Y})$ are the standard deviations of the $k^{th}$ column of $\mathbf{X}$ and the $k^{th}$ column of $\mathbf{Y}$, respectively. 
The initial value of $b_k$ is set to be $\mu_k(\mathbf{Y})-\mu_k(\mathbf{X})$, where $\mu_k(\mathbf{X})$ and $\mu_k(\mathbf{Y})$ are the averages of the $k^{th}$ column of $\mathbf{X} $ and the $k^{th}$ column of $\mathbf{Y}$,  respectively. The initial value of $\Pi$ is the matrix of size $m\times n $ with all entries equal to $\frac{1}{mn}$.
If the sample sizes $m$ and $n$ are small, we can follow the step $2(c)$, i.e., the weighted linear regression, using $m\times n$ pairs of $(X_i,Y_j)$ and weights represented by $\Pi_{i,j}^t$, and in this case we  set $\gamma_t =\frac{0.5}{t+1}$. 
However, in the case of large sample sizes, the computational cost of weighted linear regression becomes expensive,  both  in time complexity and space complexity,
making it impractical to use. Instead, we propose a stochastic linear regression in steps $2(c)$ by $2(c^\prime)$. The stochastic linear regression involves two parts. 
First, it randomly samples $B$ pairs of $(i,j)$ from the joint distribution represented by the coupling matrix. Second, it fits simple linear regression on the sampled pairs.  In this case, we have a smaller learning rate $\gamma_t =\frac{0.05}{t+1}$ since this stochastic approximation will introduce some random noises due to sampling. 
The stochastic linear regression  is more efficient and scalable for handling large datasets than fitting weighted linear regression. To be more specific, solving the weighted linear regression has time complexity $O((mn)^2)$ and space complexity $O(mn)$, while the stochastic linear regression has time complexity $O(B^2)$ and and space complexity $O(B)$.
The parameters $\lambda_1$ and $\lambda_2$ are selected by cross-validation. In practice, we recommend selecting $\lambda_1$ and $\lambda_2$ within the range of  $[\frac{0.02}{mn}\times \sum_{ij}C^0_{i,j} , \frac{0.1}{mn} \times \sum_{ij}C^0_{i,j} ]$. For the \textit{Drosophila} embryo data with $m=3,039$ and $n=1,297$ as described in Section \ref{sec:dro}
, we applied the weighted linear regression in step $2(c)$. In contrast, for the mouse cerebellum data with $m=11,626$ and $n=26,139$ in Section \ref{sec:mouse_cere}, we opted for the stochastic linear regression in $2(c^\prime)$ to avoid slow computation and potential out-of-memory issues associated with the weighted linear regression.

\subsection{Solving two tasks}
We herein introduce how to solve the two tasks shown in  \textbf{Fig.} \ref{Fig.main}b-c. 
The first task is to reconstruct spatial gene expressions, as shown in \textbf{Fig.} \ref{Fig.main}b. Given a non-marker gene in the scRNA-seq dataset, which is represented by the $g^{th}$ column of the $\mathcal{Y}$ matrix, i.e., $(y_{1g},\cdots,y_{ng})$.
The spatial expression of this gene is predicted as follows:
\begin{equation}\label{Predict_gene}
    \hat{x}_{i,g}= \frac{1}{\sum_{j=1}^n \hat{\Pi}_{i,j}}\sum_{j=1}^n \hat{\Pi}_{i,j}y_{jg}, {\rm ~for~} i=1,\cdots,m,
\end{equation}
where $\hat{x}_{i,g}$ is the predicted expression level of this non-marker gene at spot  $i$. 

The second task is to infer possible locations of single cells, i.e., the task given in \textbf{Fig.} \ref{Fig.main}c.  The probability that the  single cell $j$ is in the spot $i$ is given by 
\begin{equation}\label{Infer_loc}
    \hat{P}(\text{the cell}~ j~ \text{in the spot}~ i)= \frac{1}{\sum_{i=1}^m \hat{\Pi}_{i,j}}\hat{\Pi}_{i,j},{\rm ~for~} j=1,\cdots,n
\end{equation}

\section{Real Data Analysis}\label{sec:res}
We present the analysis of  \textit{Drosophila} embryo data and mouse cerebellum data in Sections \ref{sec:dro} and \ref{sec:mouse_cere}, respectively, while the analysis of  mouse hippocampus data,  human dorsolateral prefrontal cortex  data, and mouse visual cortex
data is provided  in the supplementary materials .

\subsection{\textit{Drosophila} embryo data}
\label{sec:dro}
\begin{figure}
\centering  
\includegraphics[width=0.9\textwidth]{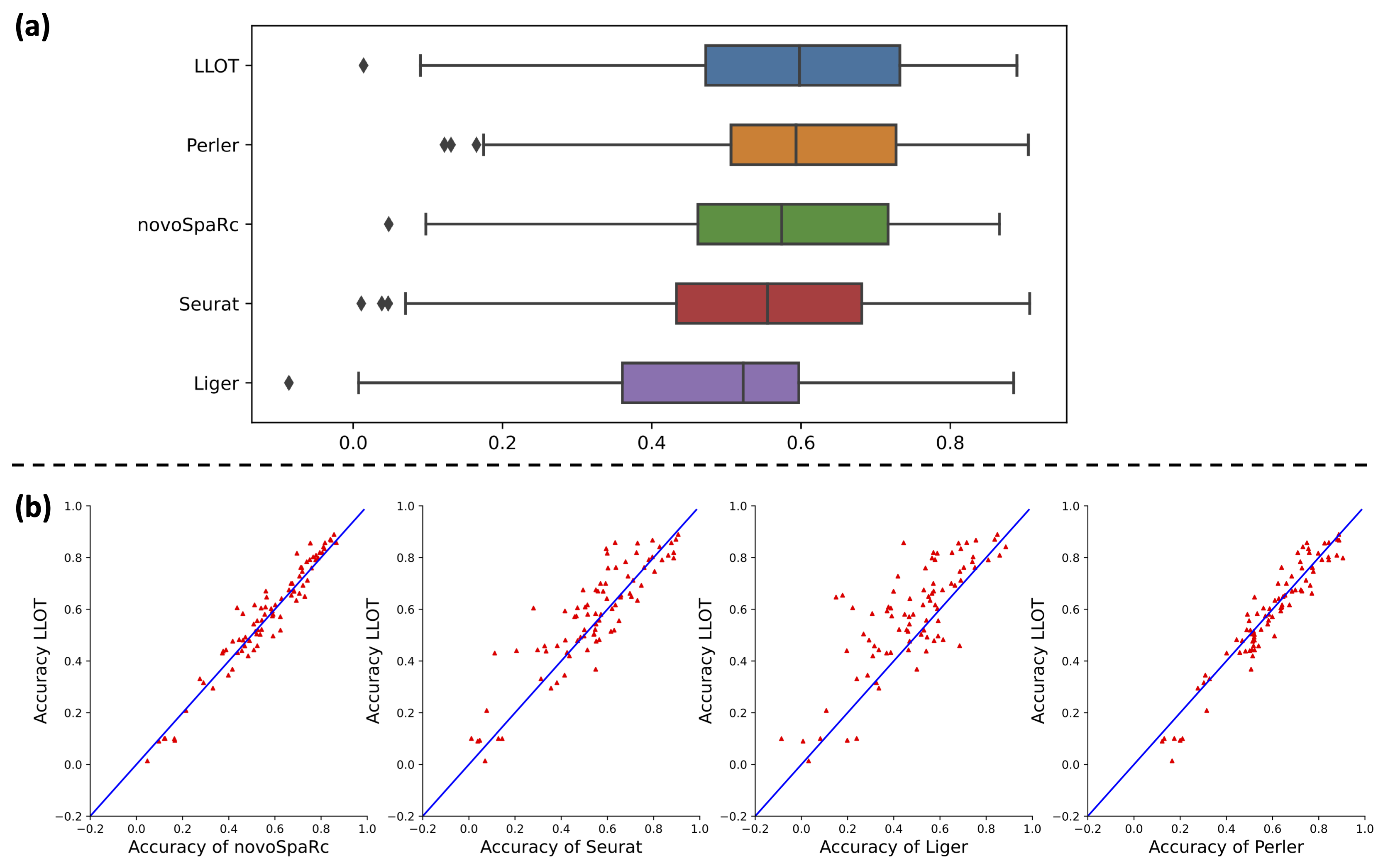}
\caption{\textbf{LOOCV benchmark on \textit{Drosophila} embryo data.}   We removed selected marker genes and evaluated the reconstruction accuracy of the removed marker gene. 
$\textbf{(a):}$ Boxplots comparing prediction accuracy of methods, as measured by Pearson Correlation Coefficients (PCC).  The accuracy is evaluated by PCC.  LLOT has the highest accuracy with the median PCC$=0.598$, followed by  Perler with optimized parameters (median PCC$=0.593$),    novoSpaRc (median PCC=$0.574$),
Seurat  (median PCC=$0.555$), Perler with default parameters (median PCC=$0.530$),  and Liger (median PCC=$0.449$). 
$\textbf{(b):}$ Scatter plots for the prediction accuracy comparison. 
The accuracy of LLOT is median PCC =$0.598$, slightly larger than  Perler  (median PCC=$0.593$) and  and 
significantly larger than novoSpaRC (median PCC =$0.574$, Wilcoxon test with p-value $0.039$),
Seurat  (median PCC =$0.555$,  Wilcoxon test with p-value $0.004$),   Wilcoxon test with p-value $3.8\times 10^{-5}$) and Liger (median  PCC =$0.449$, Wilcoxon test with p-value $1.5\times 10^{-8}$). }
\label{Drosophila}
\end{figure}

To evaluate the performance of LLOT, we applied it to a \textit{Drosophila} embryo  ISH dataset (with $p=84$ marker genes) and a  relevant scRNA-seq dataset (with $q=8,924$ genes) \citep{Distmap} and reconstructed spatial expression patterns in the embryo, as shown as the task in \textbf{Fig.} \ref{Fig.main}b.
We next conducted leave-one-out cross-validation (LOOCV) experiments to evaluate the performance of the model. In the LOOCV test, we withheld one gene, used the remaining 83 marker genes to train the model,  predicted the spatial expression of the withheld gene, and compared the prediction with the observation in spots. We repeated this experiment 84 times to test on each of the 84 marker genes,  and plotted the accuracy  distribution (\textbf{Fig.} \ref{Drosophila}). For each spot, the total expression level of  a marker gene can be  calculated as the sum of 
expression level of that gene in individual cells (previously known in the scRNA-seq dataset),
 weighted by the predicted probability that specific cell is located in the spot, i.e.,  $\text{the weight of the cell $j$ in the spot $i$ }$ is $(\Pi_{i,j}/\Sigma_k \Pi_{i,k} )$. We compared the performance of our LLOT method to Perler \citep{okochi2021model}, novoSpaRc \citep{novaspa}, 
Seurat  \citep{stuart2019comprehensive}, and Liger \citep{liger}. Here we used Pearson Correlation Coefficients as a metric of reconstruction accuracy, which is also used in a benchmark study \citep{li2022benchmarking}. We denote $\hat{x}_{i,g}(f)$ the predicted value of the $g^{th}$ gene expression in spot $i$ given by method $f$, the PCC between the prediction and the ground truth is given by 
\begin{equation}\label{pcc}
    \text{PCC}_{g}(f)=\sum_{i=1}^m (\hat{x}_{i,g}(f) \cdot x_{i,g})\bigg/\sqrt{\sum_{i=1}^m \hat{x}_{i,g}^2 (f)\cdot\sum_{i=1}^m x_{i,g}^2}.
\end{equation}
In the PCC formula presented above, we have omitted the centering step for the sake of simplicity in notation; however, the centering procedure was implemented in practice.

We show in \textbf{Fig.} \ref{Drosophila}a and \ref{Drosophila}b  the range of prediction accuracy for all compared methods. 
The performance of LLOT (median PCC =$0.598$) is slightly better than Perler (median PCC=$0.593$ after  optimized hyperparameters),  and is significantly
larger than novoSpaRC (median PCC =$0.574$, one-side Wilcoxon test p-value = $0.013$), 
Seurat  (median accuracy =$0.555$,  Wilcoxon test  p-value $0.004$), and Liger (median accuracy =$0.449$, Wilcoxon test  p-value $5\times 10^{-10}$). The null hypothesis of the Wilcoxon is  the median PCC given by LLOT is  same as median PCC given by another compared  method. The test is taken over sample of pairs $\{\big(\text{PCC}_g(\text{LLOT}\big),~(\text{PCC}_g(f)),~g=1,\cdots,84\}$, where $\text{PCC}_g(f)$ is  given by Eq.\eqref{pcc}, with $f=\text{novoSpaRc, Liger, Seurat, Perler}$. While Perler demonstrates performance comparable to LLOT, it encounters computational challenges with large-scale datasets. This is due to its reliance on high-precision density evaluations for each data point in high-dimensional spaces. In addition, we have shown that Perler performs poorly in synthetic datasets generated by scDesign3 in Section \ref{sec:simulation}.

\begin{figure}
\centering  
\includegraphics[width=0.95\textwidth]{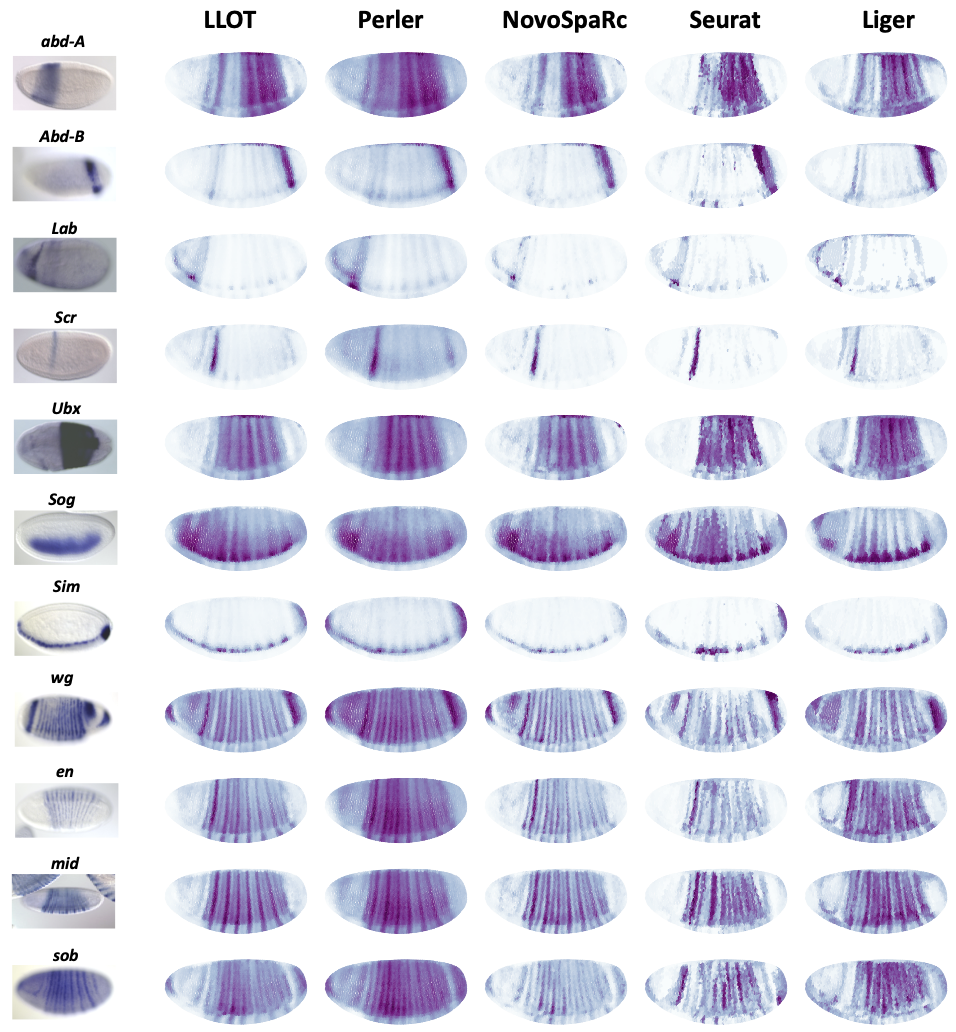}
\caption{\textbf{Prediction of non-marker genes.} We compared  predicted spatial expressions  for non-marker genes selected in  \cite{okochi2021model}. The experimental images were downloaded from BDGP (Berkeley \textit{Drosophila} Genome Project) \citep{BDGP}. 
}
\label{nonmarker}
\end{figure}

We next explored whether LLOT can also be used to predict the spatial expression of non-marker genes, i.e., those genes without  available quantitative  spatial expressions  data. 
A major obstacle to such comparison is that, currently, most of the \textit{Drosophila} embryo spatial expressions of non-marker genes exist in the form of images and have not been precisely quantified into quantitative numerical values \citep{BDGP}.  To circumvent such limitations,  we visually compared  experimental images of  Stages 4-6 \textit{Drosophila} embryos extracted from the Berkeley \textit{Drosophila} Genome Project  \citep{BDGP} with our predictions.  As seen in \textbf{Fig.} \ref{nonmarker},  LLOT is consistently among the better methods for these non-marker genes.
A recently published study \citep{okochi2021model} showed that Perler with optimized hyper-parameters for the \textit{Drosophila} embbryo data outperformed Seurat and Liger on the same set of non-marker genes. Our results show that LLOT has comparable performances.

\begin{figure}
\centering  
\subfigure[ \cite{eLife}]{
\label{eveoddref}
\includegraphics[width=0.45\textwidth]{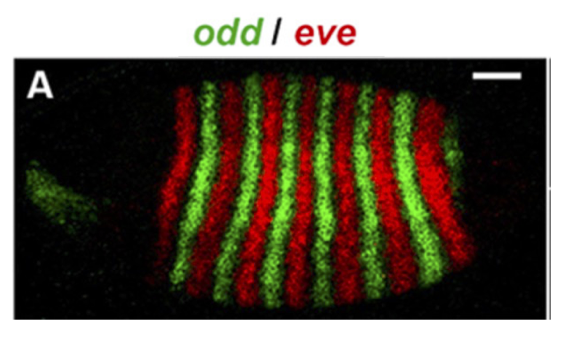}}
\subfigure[]{
\label{eveoddrec}
\includegraphics[width=0.45\textwidth]{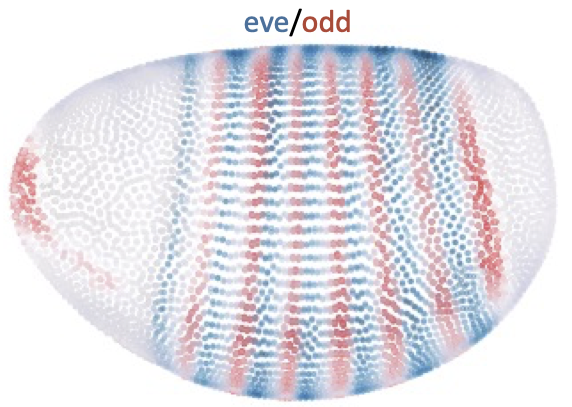}
}
\caption{\textbf{(a):} Experimental image of spatial expressions of two pair-rule genes in the \textit{Drosophila} embryo, \textit{odd} and \textit{eve}, in the study by  \cite{eLife}. 
\textbf{(b):} LLOT consistently reconstructed the non-overlapped seven-stripe patterns of \textit{odd} and \textit{eve} in the \textit{Drosophila} embryo.  }
\end{figure}

A major application of these prediction tools is that it allows us to predict spatial expressions of genes involved in the segmentation in \textit{Drosophila} embryo \citep{okochi2021model}. Segmentation in \textit{Drosophila} embryo refers to the developmental process that subdivides the embryo into a series of repeating segments along the anterior-posterior (AP) axis. This segmentation is  determined by a complex network of genes and their encoded transcription factors, which are expressed in a spatially and temporally regulated manner \citep{pbio}. 
The pair-rule genes (including \textit{eve, ftz, odd, run, h} in the set of marker genes) and segmentation polarity genes (including \textit{en, wg} in the set of non-marker genes) are  key components of this network, and are responsible for dividing the embryo into repeating segments \cite{embo,eLife,pbio}. 
We used LLOT to predict the spatial expression of these pair-rule and polarity genes. The prediction performance on \textit{eve} and \textit{odd} are shown in \textbf{Fig.} \ref{eveoddref} and \ref{eveoddrec}. \textbf{Fig.} \ref{eveoddref} is an image taken from  \cite{pbio}, showing eve and odd have high expression in distinct, non-overlapping stripes. 
\textbf{Fig.}\ref{eveoddrec}  shows that  the prediction from LLOT successfully captured the spatial pattern of these two genes.  We also compared the performance of different tools on the prediction of two segmentation polarity genes, \textit{en} and \textit{wg}, which are non-marker genes and shown in \textbf{Fig.}\ref{nonmarker}. The experimental images from BDGP shows both \textit{en} and  \textit{wg} have the 14-stripe pattern, which was precisely captured by, LLOT, Perler, and novoSpaRc, but only partially captured by  Seurat and Liger.

We next evaluated how well LLOT can infer the spatial location of cells in the \textit{Drosophila} embryo scRNA-seq dataset. This can be achieved by transforming the corresponding columns in the coupling matrix $\Pi$ as seen in \textbf{Fig.} \ref{Fig.main}(a), in which the element $\Pi_{i,j}$ represents the probability that the spot $i$ contains the cell $j$.
 This transformation involves re-normalizing columns into probabilistic vectors representing the likelihood that the cell originated from that particular spot. 
The mathematical formulation is given in  Section \ref{sec:method}.
\begin{figure}[t]
\centering  
\subfigure[]{
\label{Loc_fig}
\includegraphics[width=0.45\textwidth]{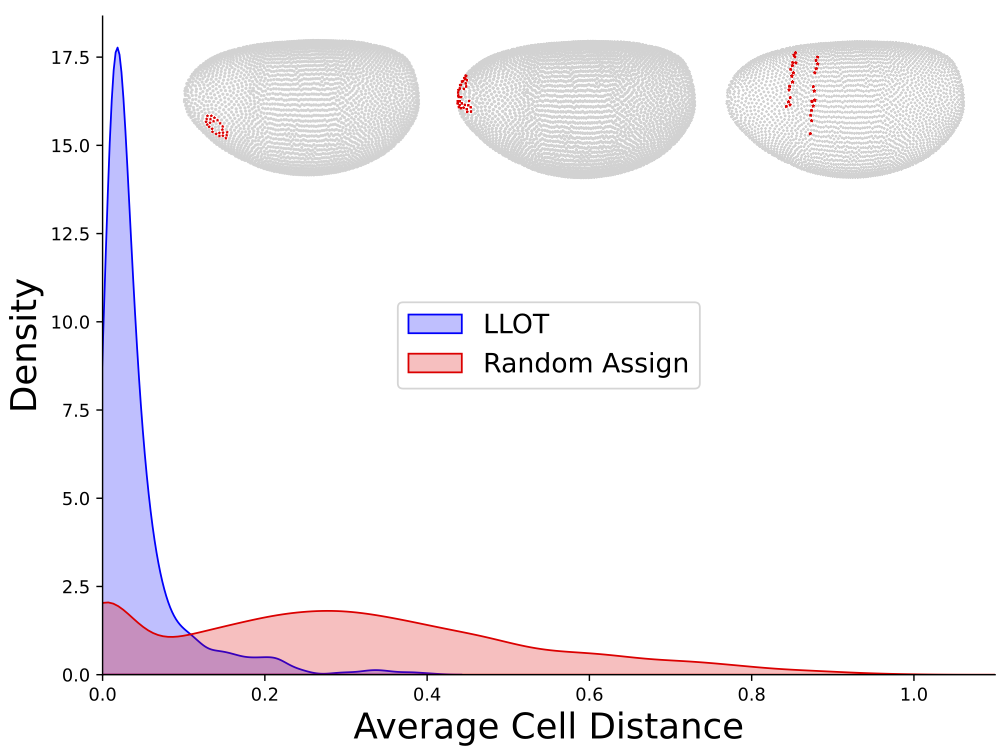}}
\subfigure[]{
\label{Cluster_loc}
\includegraphics[width=0.5\textwidth]{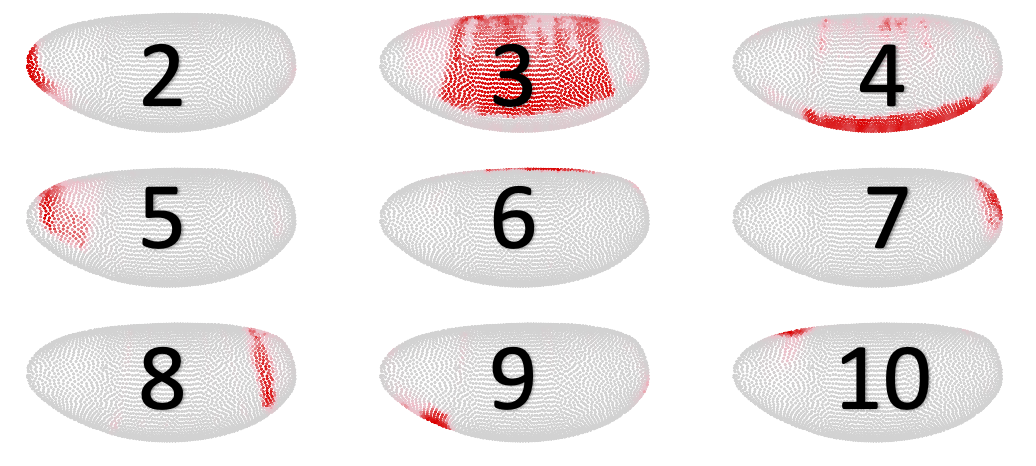}
}
\caption{\textbf{(a): Mapping cells to possible locations.} A density plot is used to illustrate the degree of specificity with which cells were assigned to possible locations. Specifically, the plot depicts the mean Euclidean distance between each cell's most probable location and the next six potential locations. A smaller average distance indicates that cells' locations are assigned with greater specificity.
Inset: Examples of inferred possible locations (labeled as red) of each single cell. 
\textbf{(b):} Each of the nine major cell clusters is assigned to a specific contiguous region in the embryo.}
\end{figure}
In our study, we randomly selected three cells from the scRNA-seq dataset and inferred their possible locations, as illustrated in $\textbf{Fig.}$     \ref{Loc_fig} (inset). The visualization results indicate that the potential locations for cells from the scRNA-seq dataset are concentrated within certain small regions.  In $\textbf{Fig.}$  \ref{Loc_fig} we used 
a density distribution similar to the study in  \cite{Distmap}, to demonstrate that our LLOT model provides 
a significantly more precise mapping of cells to tissue regions than a random assignment of single-cell origins.
As can be seen in $\textbf{Fig.}$  \ref{Loc_fig}, the random assignment (red) yields a spatial distribution of cells with large dispersion, while the spatial distribution inferred by LLOT (blue) is a more concentrated distribution. 

We next demonstrate that  LLOT can effectively reconstruct recognized cell cluster localization patterns within the \textit{Drosophila} embryo. 
The \textit{Drosophila} embryo scRNA-seq data could be clustered into eleven distinct cell types based on transcriptomic similarities \citep{Distmap}; nine cell types are major clusters in this scRNA-seq data. 
A cell type is defined in the scRNA-seq dataset and represents a subset of cells with similar gene expression profiles. 
By merging the predicted locations of cells for these nine major cell clusters, 
we reproduced results consistent with the results presented by  \cite{Distmap} (Fig. 3  in the original publication). As shown in $\textbf{Fig.}$  \ref{Cluster_loc},  each cell cluster occupies a contiguous region and represents regionally confined developmental fates \cite{hartenstein1985fate, chyb2013atlas}. The study by  \cite{Distmap} also used GO enrichment analysis to support the results and provided biological interpretations of the cell-cluster distribution results presented in $\textbf{Fig.}$  \ref{Cluster_loc}. For example, 
cluster 9, located in the anteroventral regions, is associated with future formation of the esophagus and pharynx; clusters 2 and 7 are destined to develop into the anterior and posterior midgut following the process of invagination \cite{Distmap}.
However, due to lack of ground truth, the performance of our method cannot be quantitatively evaluated at the present time.

\begin{figure}
\centering  
\includegraphics[width=0.95\textwidth]{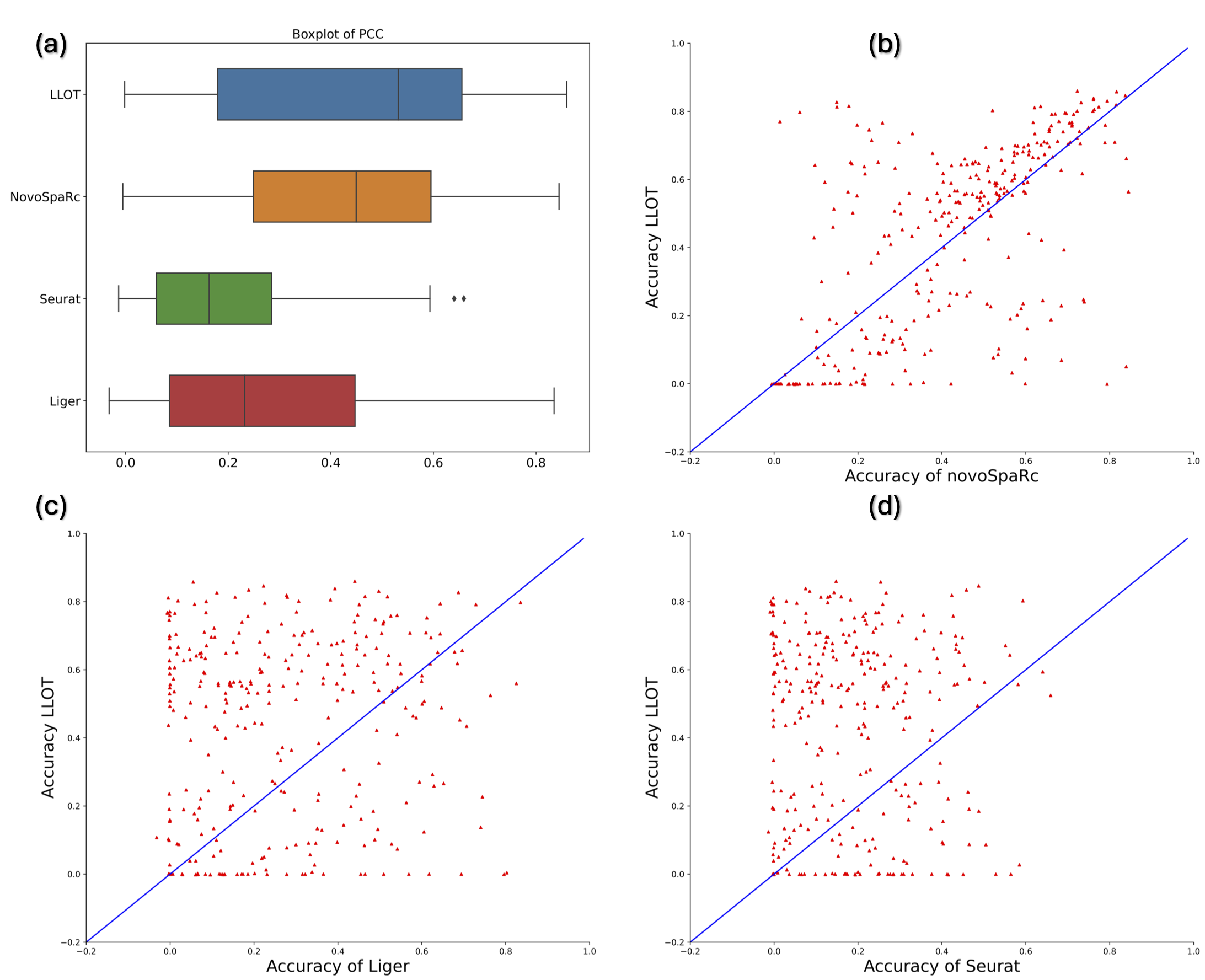}
\caption{\textbf{Boxplots and scatter plots for 10-fold CV benchmark study on mouse cerebellum data.} \textbf{(a):} Boxplots for prediction accuracy, evaluated by PCC.  
 LLOT has the highest accuracy with median PCC$=0.536$, followed by novoSpaRc (median PCC=$0.454$), Liger (median PCC=$0.232$) and 
Seurat  (median PCC=$0.163$) . \textbf{(b)-(d):} Scatter plots for prediction accuracy.
The 10-fold CV median PCC for our LLOT is $0.536$, significantly larger than novoSpaRc  (median PCC $0.454$,  Wilcoxon test: p-value=$0.019$),
Seurat  (median PCC $0.163$,  Wilcoxon test: $\text{p-value}=2.3\times 1.1^{-30}$), and Liger (median PCC $0.233$, Wilcoxon test: $\text{p-value}=6.1\times 10^{-15}$).
}
\label{Barplot2}
\end{figure}

\subsection{ Benchmark on mouse cerebellum data.}\label{sec:mouse_cere}
We next applied LLOT to  mouse cerebellum data, which was also included in the previous benchmark study by  \cite{li2022benchmarking}. The benchmark data includes the Slide-seq dataset from  \cite{cable2022robust} and a scRNA-seq dataset from  \cite{saunders2018molecular}. The Slide-seq  dataset contains  spatial gene expression data of $m=11,626$ spots, where $p=23,096$ genes are measured in each spot. 
The scRNA-seq dataset has gene expression profiles of $n=26,139$ individual cells, measuring $q=24,247$ genes and further grouped into $11$ cell clusters according to their gene expression profiles. As noted by the study of \cite{cable2022robust}, there exist significant platform effects between these two datasets, which were quantified as 
the $\log$ ratio of average gene expressions between platforms.  In addition, the cerebellum also has a complex topological structure, which makes it further challenging for spatial reconstruction.

We  showed the 
10-fold cross-validation (10-fold CV) benchmarking results evaluated by PCC in $\textbf{Fig.}$  \ref{Barplot2}, where the definition of $PCC$ is given by Eq.\eqref{pcc}.  In the 10-fold cross-validation experiment, we withheld $10\%$ of genes from the data and used the rest genes as input to predict the spatial expressions of the withheld genes. We compared LLOT with novoSpaRc, Liger, and Seurat. Perler is not included here because it suffers from high computational cost for this large-scale dataset. Similarly to  Section \ref{sec:dro}, we performed Wilcoxon test on the PCC produced by different methods. The 10-fold CV median PCC for  LLOT is $0.536$,  significantly larger than novoSpaRc  (median PCC $0.454$, Wilcoxon test: p-value=$0.019$), 
Seurat v.3 ( median PCC $0.163$,  Wilcoxon test: $\text{p-value}=2.3\times 1.1^{-30}$), and Liger ( median PCC $0.233$,  Wilcoxon test: $\text{p-value}=6.1\times 10^{-15}$).  We also used scattering plots in \textbf{Fig.} \ref{Barplot2} b-d to compare the prediction accuracy of LLOT and other individual methods. We note that this particular mouse cerebellum dataset is known to be difficult in spatial reconstruction;   novoSpaRc, Seurat, and Liger did not perform well on this particular dataset, which was previously reported in a larger benchmark study by  \cite{li2022benchmarking} (please refer to dataset 40 in Extended Data Fig.4 in that study).
Through 10-fold CV performance benchmark study, our LLOT method on the mouse cerebellum data  demonstrates a significantly improved prediction accuracy in accounting for  platform effects and  complex tissue structure.

\section{Simulation Study}\label{sec:simulation}
In this section, we investigate the performance of LLOT through numerical simulations. We generated the simulation data using scDesign3
\citep{song2024scdesign3}, which is a simulator for scRNA-seq data and spatial transcriptomics data. The simulator scDesign3 takes real data as inputs, learns marginal distributions of gene expressions  using generalized additive model for location, scale and shape  (GAMLSS) \citep{GAMLSS} to incorporate cell-type  or spatial coordinates as  covariates,
and  learns dependency between genes via vine copulas \citep{czado2019analyzing}. 
Then it generates a synthetic data from the learned probabilistic model that can mimic the pattern of the input real data. Compared to simulating data from a pre-defined distribution, scDesign3 can generate data that better fits the real data application. 

In the simulation study, we  followed the same LOOCV scheme as  Section \ref{sec:dro}
to evaluate the performance of different methods.
We  imported the  
\textit{Drosophila} embryo  ISH dataset with gene expressions and spatial coordinates,   and scRNA-seq dataset with gene expressions and cell-type labels.
\citep{Distmap}, which was used in Section \ref{sec:dro},  into the scDesign3 to generate $100$ pairs of  synthetic spatial expression and  scRNA-seq datasets.
For the scRNA-seq dataset, since only the $84$ common genes are used in the benchmark study, we only imported  these $84$ shared genes to the scDeisgn3 to improve computational efficiency. 

For each pair of synthetic datasets, we  followed the same   withhold and prediction  pipeline in Section \ref{sec:dro} for each gene, and obtain $84$ PCC as a metric of reconstruction accuracy. We applied this LOOCV procedure for the $100$ pairs of synthetic datasets and obtained  $8400$ PCC, where $100$ PCC were produced for  each gene. We compared median PCC  for each of the  $84 $ genes over $100$ experiments, and  presented the results of  seven randomly selected  genes in Table \ref{tab_main}.
The detailed results of median PCC  for all the  $84 $ genes over $100$ experiments are shown in the supplementary materials. LLOT, Liger, novoSpaRc, Seurat, and Perler achieved the highest median PCC on $53$, $ 15$, $ 3$, $ 15$, $8$ and $ 5$ genes, respectively, among $84 $ genes. 
LLOT outperformed other four compared methods in the first task on $100$ simulated datasets.

\begin{table}[h]\label{tab_main} 
\centering
\begin{tabular}{cccccccc}
Method & Gene 3 & Gene 23 & Gene 31 & Gene 41 & Gene 44 & Gene 51 & Gene 55 \\
\hline
LLOT & \textbf{0.604} & \textbf{0.829} & 0.324 & \textbf{0.638} & 0.256 & \textbf{0.696} & \textbf{0.543} \\
novoSpaRc & 0.489 & 0.769 & \textbf{0.372} & 0.621 & 0.296 & 0.661 & 0.455 \\
Seurat & 0.416 & 0.691 & 0.318 & 0.490 & 0.244 & 0.506 & 0.344 \\
Liger & 0.411 & 0.801 & 0.305 & 0.584 & \textbf{0.349} & 0.580 & 0.412 \\
Perler & 0.231 & 0.614 & 0.132 & 0.355 & -0.045 & 0.504 & 0.236 \\
\hline \end{tabular}
\caption{Reconstruction accuracy of seven randomly selected genes evaluated by median PCC over 100 synthetic datasets.  The bold number indicates the best accuracy among five methods. }
\end{table}\vspace{0.5cm} 

\section{Discussion}\label{sec:dis}
In this paper, we describe  \textit{Laplacian Linear Optimal Transport} (LLOT), which effectively integrates spatial  transcriptomics and single-cell RNA sequencing (scRNA-seq) datasets into a coupling matrix.  We   demonstrated that LLOT can accurately predict spatial expressions of single cells and infer the locations of single cells. We showed that LLOT generally performed as well as or better than other contemporary methods.  Compared with contemporary methods, LLOT has two major advantages and innovations.  First, it incorporates a linear map to model platform effects. The linear map is interpretable as it explicitly models the platform effects. It is also
computationally efficient since the parameters can be estimated by  stochastic linear regression at each iteration with time complexity $O(B^2)$ and space complexity $O(B)$, where $B$ is the batch size, i.e., the number of cell-spot pairs in random sub-sampling during training. This provides a significant advantage when dealing with large-scale datasets from different sources. Second, it employs the novel framework of Laplacian Optimal Transport to leverage the spatial information in the spatial transcriptomics data. The Laplacian optimal transport involves a Laplacian regularization as a local smoother which can capture the complex  topology structure of tissues, and 
generate a probabilistic map with local similarity and local smoothness, referred to as the coupling matrix $\Pi$ between single cells and spots; the element $\Pi_{i,j}$ of the coupling matrix represents the likelihood that the spot $i$ contains the cell $j$.

We benchmarked LLOT with several other state-of-the-art computational methods on the task of predicting  spatial gene expressions on three datasets, including  \textit{Drosophila} embryo, mouse cerebellum, mouse hippocampus data and synthetic datasets generated by scDesign3.  In summary, LLOT either outperformed other methods  or achieved similar results in spatial expression predictions. We then demonstrated LLOT can infer segmentation-related spatial expressions and infer 
locations of cells in \textit{Drosophila} embryo in a biologically meaningful way.
We also extended LLOT to  layer recovery in 
human dorsolateral prefrontal cortex data under two scenarios proposed in  CeLEry \citep{zhang2023celery} and to perform cell-type deconvolution, associated with benchmark study. These results showed that LLOT is an effective and  reliable tool in the realms of spatial transcription reconstruction, cell-location recovery and cell-type deconvolution.


Our LLOT method also has several limitations and room for future improvement. First, 
a key computational challenge lies in the approximation of optimal coupling matrices in large-scale datasets, which can be time-consuming since the computational complexity of solving an optimal transport problem in our method is $O((m+n)^2/\epsilon^2)$ where  $\epsilon$ is the approximation accuracy \cite{dvurechensky2018computational}. This issue may be addressed by incorporating the recent work on stochastic approximation in optimal transport proposed by  \cite{genevay2016stochastic} and   \cite{ballu2020stochastic}; although it is important to acknowledge this may result in a compromise on approximation accuracy. Thus, more research is needed in this area.
 Second, 
while LLOT  has demonstrated better performance than two other optimal-transport-based methods, it did not perform as well as RCTD in the cell-type deconvolution task \citep{cable2022robust}. This is especially noticeable when the cell-type compositions in the spatial transcriptomics dataset and scRNA-seq dataset differ significantly. The incorporation of unbalanced optimal transport \citep{chizat2018scaling} to introduce cell-type-specific soft constraints on the marginal distribution may be a potential solution to this challenge.
Third, we assumed that platform effects are variable among genes but invariant across cell types  for the same gene.   This can potentially be improved by assigning platform effect parameters to single cells based on their specific cell type labels. This would allow for a more accurate correction of platform effects. However, special consideration is needed to prevent over-fitting. Fourth, it will be interesting to modify the Laplacian regularization in LLOT to incorporate boundary information from  histology images of tissues or other additional information, such as temporal information.

In summary, LLOT provides a robust, interpretable framework to integrate  spatial transcriptomics and scRNA-seq data.  By effectively correcting platform effects and incorporating spatial information, we show that LLOT can provide a more accurate reconstruction of spatial gene expressions. As we continue to  refine this method, we anticipate that it will contribute significantly to enhancing our understanding of complex tissue structures and functions.

\bibliographystyle{agsm}

\bibliography{referenceclean}
\end{document}